\documentstyle[preprint,pra,aps,eqsecnum,amsfonts]{revtex}

\tightenlines

\begin{document}

\draft
\title{A simple expression for the terms in the Baker-Campbell-Hausdorff
series}
\author{Matthias W. Reinsch}
\address{Department of Physics, University of California, Berkeley, California
94720, {\tt reinsch@uclink4.berkeley.edu}}
\date{\today}
\maketitle

\begin{abstract}

A simple expression is derived for the terms in the
Baker-Campbell-Hausdorff series.  One formulation of the result
involves a finite number of operations with matrices of rational
numbers.  Generalizations are discussed.

\end{abstract}


\section{Introduction}

The Baker-Campbell-Hausdorff series has a long history and has
applications in a wide variety of problems, as explained in
Refs.~\onlinecite
{Goldberg,Kobayashi,Oteo,NT,Dynkin,T,Bose,Reutenauer,WM,Gilmore,Magnus1,Magnus2,MKO}.
In a classic paper,
Goldberg\cite{Goldberg} was able to derive an integral expression for the
coefficients in the general term, and this result is still used
today\cite{Kobayashi,Oteo,NT} to calculate the Baker-Campbell-Hausdorff series.

In this paper, we present a simple method for calculating the terms in
the Baker-Campbell-Hausdorff series.  The process can be carried out by
hand, and it is easily implemented on a computer.

\section{Statement of theorem}
\label{thm}

We let $z = \log( e^x \, e^y)$ denote the Baker-Campbell-Hausdorff
series for noncommuting variables $x$ and $y$.  Our result for the
$n$-th order term $z_n$ in this series is given by the following
procedure, which involves only a finite number of matrix
multiplications.  We state our results without reference to
commutators.  If an expression in terms of commutators is desired, our
expression can be transformed using the substitution due to
Dynkin\cite{Oteo,Dynkin,T,Bose}.  Each product involving $x$ and $y$
variables is replaced by $\frac{1}{n}$ times the corresponding
iterated commutator of the same sequence of $x$'s and $y$'s.  The quantity
$z_n$ is invariant under this transformation.

To calculate $z_n$, the entire $n$-th order term in the
Baker-Campbell-Hausdorff series, we compute a certain polynomial in
$n$ (ordinary commuting)
variables, $\sigma_1, \ldots, \sigma_n$, and then make a
replacement, as described below.  We begin by defining two
$(n+1)\times(n+1)$ matrices $F$ and $G$ by
\begin{equation}
F_{ij} = \frac{1}{(j-i)!}
\label{f}
\end{equation}
and
\begin{equation}
G_{ij} = \frac{1}{(j-i)!} \, \prod_{k=i}^{j-1} \, \sigma_k \, .
\label{g}
\end{equation}
These equations are valid for all $i$ and $j$ from $1$ to $n+1$, with
the usual convention that the reciprocal of the factorial of a
negative integer is zero.  Written out explicitly, the matrices are
\begin{equation}
F = \left(\begin{array}{ccccccc} 1 \; {} & 1 & \frac{1}{2} & \frac{1}{6} &
\ldots & {} & {} \\
{} & 1 & 1 & \frac{1}{2} & \frac{1}{6} & \ldots & {} \\
{} & {} & 1 & 1 & \frac{1}{2} & \frac{1}{6} & \ldots \\
{} & {} & {} & . & {} & {} & {} \\
{} & {} & {} & {} & . & {} & {} \\
{} & {} & {} & {} & {} & . & {} \\
{} & {} & {} & {} & {} & {} & 1 \end{array}\right)
\end{equation}
and
\begin{equation}
G = \left(\begin{array}{ccccccc} 1 \; {} & \sigma_1 & \frac{1}{2}\sigma_1
\sigma_2 & \ldots & {} & {} & {} \\
{} & 1 & \sigma_2 & \frac{1}{2}\sigma_2 \sigma_3  & \ldots & {} & {} \\
{} & {} & 1 & \sigma_3 & \frac{1}{2} \sigma_3 \sigma_4  & \ldots {} & \\
{} & {} & {} & . & {} & {} & {} \\
{} & {} & {} & {} & . & {} & {} \\
{} & {} & {} & {} & {} & . & \sigma_n \\
{} & {} & {} & {} & {} & {} & 1 \end{array}\right) \, .
\end{equation}
Although it is not necessary for the calculation of results, we point
out at this point that the matrices $F$ and $G$ are exponentials of
very simple matrices.  We define two $(n+1)\times(n+1)$ matrices $M$
and $N$ by
\begin{equation}
M_{ij} = \delta_{i+1,j}
\label{m}
\end{equation}
and
\begin{equation}
N_{ij} = \delta_{i+1,j} \, \sigma_i \, .
\label{n}
\end{equation}
These equations are valid for $i$ and $j$ ranging from $1$ to $n+1$.  A simple
application of the definition of the exponential function gives $F =
\exp M$ and $G = \exp N$.  Written out explicitly, these statements
are
\begin{equation}
F = \exp \left(\begin{array}{ccccccc}
0  & 1 & 0 & \ldots & {} & {} & {} \\
{} & 0 & 1 & 0 &  \ldots & {} & {} \\
{} & {} & . & {} & {} & {} & {} \\
{} & {} & {} & . & {} & {} & {} \\
{} & {} & {} & {} & . & {} & {} \\
{} & {} & {} & {} & {} & 0 & 1 \\
{} & {} & {} & {} & {} & {} & 0 \end{array}\right)
\label{FexpM}
\end{equation}
and
\begin{equation}
G = \exp \left(\begin{array}{ccccccc}
0  & \sigma_1 & 0 & \ldots & {} & {} & {} \\
{} & 0 & \sigma_2 & 0 &  \ldots & {} & {} \\
{} & {} & . & {} & {} & {} & {} \\
{} & {} & {} & . & {} & {} & {} \\
{} & {} & {} & {} & . & {} & {} \\
{} & {} & {} & {} & {} & 0 & \sigma_n \\
{} & {} & {} & {} & {} & {} & 0 \end{array}\right) \, .
\label{GexpN}
\end{equation}
The matrices $M$ and $N$ will be used later in a proof.

Our expression for the $n$-th order term in the
Baker-Campbell-Hausdorff series is
\begin{equation}
z_n = T \, (\log FG)_{1, n+1} \, .
\label{a}
\end{equation}
The indices on the right-hand side of this equation indicate the
upper-right element of the matrix $\log FG$.  The operator $T$ 
replaces products of $\sigma$-variables with products of $x$ and $y$
according to the following procedure.
The polynomial $(\log FG)_{1, n+1}$ is a sum of terms, each of which may be
written as a rational number times $\sigma_1^{\mu_1} \,
\sigma_2^{\mu_2} \, \cdots \, \sigma_n^{\mu_n}$, where the $\mu_i$ are
either $0$ or $1$ (no exponents greater than 1 occur, as explained
later in this paper).  Next, $\sigma_i^{\mu_i}$ is replaced with $x$ if
$\mu_i = 0$ and $y$ if $\mu_i = 1$.  Thus each $\sigma_i$ that occurs
(to the first power) in a term indicates that a $y$ is to be placed at
the $i$-th location in the product of $x$ and $y$ variables.  For
example, in the case $n = 6$, we have $T(\sigma_2 \, \sigma_4 \, \sigma_5)
= xyxyyx$.
The operator $T$ is a vector-space isomorphism from the space of
polynomials in the $\sigma$-variables (with $\mu_i \le 1$) to the space of
linear combinations of products that have $n$ factors that are either $x$
or $y$.

The $\log$ operation in Eq.~(\ref{a}) is simple because $FG$ is equal
to the $ (n+1)\times(n+1)$ identity matrix (which we denote by $I$)
plus a matrix that is strictly upper triangular.
Thus the series for $\log[I+(FG-I)]$ terminates after finitely many terms.
\begin{equation}
\log FG = - \sum_{q = 1}^n \frac{(-1)^q}{q} \, (FG-I)^q \, .
\end{equation}
The calculation of $z_n$, the order $n$ term
in the Baker-Campbell-Hausdorff series, can therefore be carried
out with a finite number of simple operations.  There are no sums over
partitions, operations with noncommuting variables, translations of binary
sequences into descriptions in terms of block lengths, etc.

\section{Examples}

Let us begin by working out the example of $n=1$.  We have
\begin{equation}
F = \left(\begin{array}{cc} 1 & 1 \\
0 & 1 \\ \end{array}\right)
\end{equation}
and
\begin{equation}
G = \left(\begin{array}{cc} 1 & \sigma_1 \\
0 & 1 \\ \end{array}\right) \, .
\end{equation}
From this follows
\begin{equation}
FG = \left(\begin{array}{cc} 1 & 1+\sigma_1 \\
0 & 1 \\ \end{array}\right)
\end{equation}
and
\begin{eqnarray}
z_1 &=& T \, (\log FG)_{1, 1+1} = T \, (\sigma_1^0+\sigma_1^1) \\
&=& x+y \, .
\end{eqnarray}

Next let us work out the example of $n=2$.  We have
\begin{equation}
F = \left(\begin{array}{ccc} 1 & 1 &\frac{1}{2} \\
0 & 1 & 1\\ 0 & 0 & 1 \end{array}\right)
\end{equation}
and
\begin{equation}
G = \left(\begin{array}{ccc} 1 & \sigma_1 & \frac{1}{2} \sigma_1 \sigma_2 \\
0 & 1 & \sigma_2 \\ 0 & 0 & 1 \end{array}\right) \, .
\end{equation}
From this follows
\begin{equation}
FG = \left(\begin{array}{ccc} 1 & 1+\sigma_1 & \frac{1}{2} + \sigma_2 +
\frac{\sigma_1 \, \sigma_2}{2} \\
0 & 1 & 1+\sigma_2 \\ 0 & 0 & 1\\ \end{array}\right)
\end{equation}
and
\begin{equation}
(FG-I)^2 = \left(\begin{array}{ccc} 0 & 0 & 1 + \sigma_1 + \sigma_2 +
\sigma_1 \, \sigma_2 \\
0 & 0 & 0 \\ 0 & 0 & 0\\ \end{array}\right) \, ,
\end{equation}
so that
\begin{eqnarray}
z_2 &=& T \, (\log FG)_{1, 2+1} =
T \left( \frac{1}{2} \sigma_1^0 \, \sigma_2^1 - \frac{1}{2} \sigma_1^1 \,
\sigma_2^0 \right)
\label{n2}
\\
&=& \frac{1}{2}(xy-yx) \, .
\end{eqnarray}

For the case $n=3$, the equations result in
\begin{eqnarray}
z_3 &=& T\left(\frac{1}{12} \sigma_1 - \frac{1}{6} \sigma_2 +
\frac{1}{12} \sigma_3 + \frac{1}{12} \sigma_1 \, \sigma_2
- \frac{1}{6} \sigma_1 \, \sigma_3 + \frac{1}{12}
\sigma_2 \, \sigma_3 \right)\\
&=&
\frac{1}{12} yxx - \frac{1}{6} xyx + \frac{1}{12} xxy  + \frac{1}{12} yyx
- \frac{1}{6} yxy + \frac{1}{12} xyy \, .
\end{eqnarray}

The case $n=4$ works out to be
\begin{eqnarray}
z_4 &=& T \left(
-\frac{1}{24} \sigma_1 \, \sigma_2 + \frac{1}{12} \sigma_1 \, \sigma_3 -
\frac{1}{12} \sigma_2 \, \sigma_4 + \frac{1}{24} \sigma_3 \, \sigma_4 \right)\\
&=&
-\frac{1}{24} yyxx + \frac{1}{12} yxyx - \frac{1}{12} xyxy + \frac{1}{24} xxyy
\, .
\end{eqnarray}

These results and higher-order calculations not shown here agree with
results published in the literature\cite{Oteo,NT}.  As an example of
the types of coefficients that occur, when $n$ is $7$ our formula
gives a coefficient of $-1/1512$ for the $yxxxyyy$ term, and this
agrees with the literature result.

\section{Proof of Theorem}

We begin by considering the Baker-Campbell-Hausdorff series for $\log\left(
e^M \, e^N\right)$, where the $(n+1)\times(n+1)$ matrices $M$ and $N$ are
defined in Eqs.~(\ref{m}) and (\ref{n}).  The matrices $M$ and $N$ are
written out explicitly on the right-hand sides of Eqs.~(\ref{FexpM})
and (\ref{GexpN}).  They have nonzero elements only on the first
superdiagonal.  Therefore, a product having $m$ factors that are
either $M$ or $N$ will have nonzero elements only on the $m$-th
superdiagonal.  Thus, the upper-right element of the matrix $\log\left( e^M
\, e^N\right)$ is equal to the upper-right element of the matrix that is the
order $n$ term in the Baker-Campbell-Hausdorff series for $\log\left( e^M
\, e^N\right)$.  We write this as
\begin{equation}
\left[ \log\left( e^M \, e^N\right) \right]_{1, n+1} = \sum_W \, C(W) \,
[\Pi(W)]_{1, n+1} \, ,
\label{a2}
\end{equation}
where the sum runs over all ``words'' $W$ of length $n$ (ordered $n$-tuples
of elements that are either the symbol $M$ or the symbol $N$),
$C(W)$ denotes the coefficient of $W$ in the order $n$ term in the
Baker-Campbell-Hausdorff series, and $\Pi(W)$ denotes a product of $M$
and $N$ matrices as specified by the word $W$.

We now show that $[\Pi(W)]_{1, n+1}$ is a product of $\sigma$-variables whose
indices give the positions of the $N$'s in the word $W$.
We let the matrix $\Pi(W)$ act on an $(n+1)$-component
column vector that is all zeroes except the lowest element, which is a
$1$.  After each multiplication by an $M$ or an $N$ matrix, the
location of the nonzero element in the column vector moves up by one
step.  If the matrix multiplying the column vector is an $N$, the
nonzero element in the column vector gets multiplied by a $\sigma$.
The index on the $\sigma$ gives the location of the $N$ matrix in the
word $W$, as can be seen by looking at the structure of the $N$
matrix, shown on the right-hand side of Eq.~(\ref{GexpN}).  After all
of the $n$ matrices in the word $W$ have acted on the column vector,
the nonzero element in the vector is at the top, and this element is a
product of $\sigma$-variables whose indices describe the word $W$ in
the manner explained above.  The top element of the vector obtained by
letting a matrix act on the initial column vector described above is
the upper-right element of the matrix.  Thus we have shown that $[\Pi(W)]_{1,
n+1}$ is a product of $\sigma$-variables whose indices give the
positions of the $N$'s in the word $W$.  This fact together with the
relations $F = \exp M$ and $G = \exp N$ and Eq.~(\ref{a2}) proves
Eq.~(\ref{a}).

\section{Alternative formulation}

In this section we present a result equivalent to the result presented
above, but the matrix operations involve only numbers.  We will
express the $n$-th order term, $z_n$, as a linear combination of terms
of the form $(x+\sigma_1 \, y) (x+\sigma_2 \, y) \cdots (x+\sigma_n \,
y)$, where the $\sigma$'s are either $+1$ or $-1$.  There are
$2^n$ such terms.  Our result is that the coefficient of a given term
is $2^{-n}$ times the value of the polynomial $(\log FG)_{1, n+1}$
defined Sec.~\ref{thm}, with the corresponding $\sigma$-values substituted in.
This number can be computed by substituting the $\sigma$-values into
the $G$ matrix before doing the matrix operations.  These statements
can be summarized as
\begin{equation}
z_n = 2^{-n} \, \sum_{\sigma_1, \cdots, \sigma_n} \, (\log FG)_{1, n+1} \,
(x+\sigma_1 \, y) (x+\sigma_2 \, y) \cdots (x+\sigma_n \, y) \, ,
\label{27}
\end{equation}
where the sum is over all $2^n$ possible assignments of $\pm 1$ to the
$\sigma$-variables.

Let us work out an example.  For $n=2$ the equation becomes
\begin{eqnarray}
z_2 &=& 2^{-2} \, \sum_{\sigma_1, \sigma_2} \, \left( \frac{\sigma_2}{2} -
\frac{\sigma_1}{2} \right) \, (x+\sigma_1 \, y) (x+\sigma_2 \, y) \\
&=& \frac{1}{4} \left[ \left(\frac{1}{2} - \frac{-1}{2}\right) (x - y)
(x + y) +
\left(\frac{-1}{2} - \frac{1}{2}\right) (x + y) (x - y) \right ] \\
&=& \frac{1}{2}(xy - yx) \, ,
\end{eqnarray}
where we have used Eq.~(\ref{n2}).

A calculation of $z_n$ without the use of multiplication of matrices
of polynomials would proceed in the following way.  For every choice of
$+1$ or $-1$ values for the $\sigma$-variables one computes the value
of $2^{-n} \, (\log FG)_{1, n+1}$.  This involves a finite number of
operations with numbers.  The result is the coefficient of $(x+\sigma_1 \,
y) (x+\sigma_2 \, y) \cdots (x+\sigma_n \, y)$ in the expression for
$z_n$ in Eq.~(\ref{27}).  Next, we imagine the process of expanding
all of the products $(x+\sigma_1 \, y) (x+\sigma_2 \, y) \cdots
(x+\sigma_n \, y)$.  The result of this operation
is a sum over words in the variables
$x$ and $y$.  To get the coefficient of a particular word one sums all
of the coefficients calculated from $2^{-n} \, (\log FG)_{1, n+1}$
with a sign given by the product of the $\sigma$'s at the locations of
the $y$'s in the word.  For example, if the word is $xxyyxy$ then the
coefficients are summed with signs given by $\sigma_3 \, \sigma_4 \,
\sigma_6$.

We now prove Eq.~(\ref{27}).  We consider the order $n$ term in
$\log( e^{x+y} \, e^{x-y})$.  This may be written in two ways,
\begin{equation}
\sum_{\sigma_1,\ldots,\sigma_n} \, C(\sigma_1,\ldots,\sigma_n) \, 
(x+\sigma_1 \, y) \cdots (x+\sigma_n \, y) = \sum_W \, C^\prime(W) \, W \, .
\label{two}
\end{equation}
The sum on the left-hand side is a sum over all assignments of $+1$ or $-1$
to the variables $\sigma_1,\ldots,\sigma_n$.  The coefficient
$C(\sigma_1,\ldots,\sigma_n)$ is the usual coefficient in the
Baker-Campbell-Hausdorff series, with the $\sigma$'s identifying a word.
The right-hand side of
Eq.~(\ref{two}) results from multiplying out all of the products
$(x+\sigma_1 \, y) \cdots (x+\sigma_n \, y)$.  It is a sum over words
in $x$ and $y$, and $C^\prime(W)$ denotes the resulting coefficient of
the word $W$.  [In this context we use the term ``word'' to denote an actual
product of a certain sequence of $x$ and $y$ variables, because such a product
does not evaluate to become something else, as it did in the case of the
products of $M$ and $N$ matrices in the previous section.  Thus, the notation
$\Pi(W)$ is not needed in Eq.~(\ref{two}).]
The coefficient $C^\prime(W)$ of a particular word $W$
can be expressed in terms of the $C(\sigma_1,\ldots,\sigma_n)$.  For
every set of values for $\sigma_1,\ldots,\sigma_n$ in the sum on the
left-hand side we get a contribution to $C^\prime(W)$ of
$C(\sigma_1,\ldots,\sigma_n)$ times the product of the $\sigma$'s that
correspond to the $y$'s in $W$.  This product is the same as the
product of $\sigma^{(W)}_i$, where the $\sigma^{(W)}$ values describe
the word $W$ ($\sigma^{(W)}_i$ is +1 if the $i$-th factor in $W$ is $x$,
and $\sigma^{(W)}_i$ is -1 if the $i$-th factor in $W$ is $y$),
and the product runs over $i$-values corresponding to
negative $\sigma$'s.  Thus, $C^\prime(W)$ is precisely the polynomial
in $\sigma^{(W)}$ given in the first theorem.  Now we transform from
$x$ and $y$ to new variables according to $x+y={\tilde x}$ and
$x-y={\tilde y}$.  This implies $x=({\tilde x}+{\tilde y})/2$ and
$y=({\tilde x}-{\tilde y})/2$, and the right-hand side of
Eq.~(\ref{two}) (which equals $\log e^{\tilde x} \, e^{\tilde y} $)
becomes the right-hand side of Eq.~(\ref{27}), after the tildes have
been dropped, which is justified since only variables with tildes occur
at that point.  The
$n$ factors of $1/2$ are collected into the factor of $2^{-n}$ in
Eq.~(\ref{27}).

\section{Symmetries of the coefficients}

The calculation described in the penultimate paragraph of the
preceding section involves repeated evaluation of $(\log FG)_{1, n+1}$
with values of $+1$ and $-1$ substituted in for the
$\sigma$-variables.  These numbers are the coefficients in the sum in
Eq.~(\ref{27}).  We begin this section by showing that half of the
resulting numbers will be zero because of a basic symmetry of the
Baker-Campbell-Hausdorff series.  Then we show that some of the
nonvanishing coefficients can be obtained from other ones.  These
considerations reduce the computational work involved in calculating
the coefficients in Eq.~(\ref{27}).

The relationship $e^z = e^x \, e^y$ implies $e^{-z} = e^{-y} \, e^{-x}$ and
\begin{equation}
z = - \log e^{-y} \, e^{-x} \, .
\end{equation}
From this we see that swapping the $x$'s and $y$'s in the $n$-th order
term $z_n$ gives the
same result as multiplying $z_n$ by $(-1)^{n-1}$.  In the present context,
the $\sigma$-variables are being assigned values of $+1$ or $-1$, so we
may use the relationship $\sigma_i^2 = 1$.  The preceding statement about
the symmetry of $z_n$ is equivalent to
\begin{equation}
(\log FG)_{1, n+1} \, \prod_{i=1}^n \sigma_i = (-1)^{n-1}\,(\log FG)_{1, n+1} .
\end{equation}
Multiplication by $\prod_{i=1}^n \sigma_i$ effects a swapping of $x$ and
$y$, because of the relationship $\sigma_i^2 = 1$.
If the number of $\sigma$'s that are $+1$ is even then $\prod_{i=1}^n \sigma_i$
will equal $(-1)^n$ because all of the $\sigma$'s in the product may be
replaced with $-1$ without changing the value of the product.  Therefore,
if the number of $\sigma$'s that are $+1$ is even, then $(\log FG)_{1, n+1}$
is zero.  Elementary combinatorics shows that this condition holds for one-half
of the terms in the sum in Eq.~(\ref{27}).  In the case of odd $n$
greater than 1, there is
one additional term which vanishes, namely the one for which all of the
$\sigma$'s are $+1$.  This is because the matrices $M$ and $N$ are
equal and therefore commute.  Thus $\log FG$ is equal to
$M + N$ and the upper-right element of $\log FG$ is zero.
[In the case of even $n$, $(\log FG)_{1, n+1}$ of course also vanishes
when all of the
$\sigma$'s are $+1$, but this vanishing has already been counted in
the discussion above.]

Thus far, we have identified coefficients in the sum in Eq.~(\ref{27})
that are zero for symmetry reasons.  This author has searched up through
order $n = 15$ and found all of the remaining coefficients to be nonzero.

A further symmetry of $(\log FG)_{1, n+1}$ is
\begin{equation}
[(\log FG)_{1, n+1}](\sigma_1, \ldots, \sigma_n) = (-1)^{n-1}
[(\log FG)_{1, n+1}](\sigma_n, \ldots, \sigma_1) \, .
\label{reverse}
\end{equation}
The notation on the left-hand side of this equation indicates explicitly that
$(\log FG)_{1, n+1}$ is a function of the $n$ variables
$\sigma_1, \ldots, \sigma_n$.
On the right-hand side, these $n$ quantities are inserted into the
function in the reversed order.  The fact that these two values of
the function are related by a factor of $(-1)^{n-1}$ is due to a
symmetry of the Baker-Campbell-Hausdorff series.  It follows immediately
from results in Ref.~\onlinecite{Goldberg} that the coefficient of a
word in the variables $x$ and $y$ is equal to $(-1)^{n-1}$ times
the coefficient of the word obtained by reversing the order of
the factors in the original word.  This implies that
$T \{ [ (\log FG)_{1, n+1} ] (\sigma_n, \ldots, \sigma_1) \}$ (where
$T$ is the operator defined in Sec.~\ref{thm}), which is
the order $n$ term in the Baker-Campbell-Hausdorff series with the
sequence of the factors in each term reversed, is equal to $(-1)^{n-1}$
times $T \{ [ (\log FG)_{1, n+1} ] (\sigma_1, \ldots, \sigma_n) \}$.
Because the vector-space isomorphism $T$ is invertible, this proves
Eq.~(\ref{reverse}).  This equation is useful because it can be used
to avoid carrying out unnecessary evaluations of $(\log FG)_{1, n+1}$.

\section{Computer implementation}

The methods presented in this paper can easily be used with computers.
A simple example of how the results of Sec.~\ref{thm} can be
implemented using Mathematica is shown below.
It is not necessary to load any special packages to run this code.
This example is
oriented toward ease of coding.  Faster implementations are possible.
The first program gives the polynomial to the right of the
$T$ operator in Eq.~(\ref{a}), and the
second program (for $n>1$) translates this into $z_n$, the corresponding
expression in terms of $x$ and $y$.
 
\begin{verbatim}
p[n_] := p[n] = ( F = Table[1/(j-i)!,{i,n+1},{j,n+1}];
         G = Table[1/(j-i)! Product[s[k],{k,i,j-1}],{i,n+1},{j,n+1}];
         qthpower = IdentityMatrix[n+1]; FGm1 = F.G - qthpower; Expand[
         -Sum[qthpower=qthpower.FGm1; (-1)^q / q qthpower,{q,n}][[1,n+1]]])

translated[n_] := (temp = Expand[Product[s[k]^2, {k,n}] p[n]];
         Sum[term = Apply[List, temp[[i]]]; term[[1]] Apply[StringJoin,
         Take[term,-n] /. {s[i_]^2->"x",s[i_]^3->"y"}], {i,Length[temp]}])

In[3]:= translated[4]

        xxyy   xyxy   yxyx   yyxx
Out[3]= ---- - ---- + ---- - ----
         24     12     12     24

\end{verbatim}

\section{Other Series}

As in the case of Goldberg's results\cite{Goldberg}, the methods of
this paper can be used to calculate $\log f(x) f(y)$, where $f(x)$ is
an arbitrary power series with $f(0) = 1$.  The only changes that are
necessary are that occurrences of the exponential function such as
those in Eqs.~(\ref{FexpM}) and (\ref{GexpN}) must be replaced with
the function $f$.  The matrices $M$ and $N$ have the property that
when they are raised to the power $n+1$ the result is zero, so the
calculation of $f(M)$ and $f(N)$ terminates after finitely many matrix
operations.

\section{Generalized Baker-Campbell-Hausdorff series}

The methods presented in Sec.~\ref{thm} may also be used to calculate the
terms in generalized Baker-Campbell-Hausdorff series.  For example,
if $z = \log e^x \, e^y \, e^w$ then the $n$-th order term may be found
as follows.  Matrices $F$ and $G$ are defined as in Eqs.~(\ref{f}) and
(\ref{g}), and a matrix $H$ is defined by
\begin{equation}
H_{ij} = \frac{1}{(j-i)!} \, \prod_{k=i}^{j-1} \, \tau_k \, ,
\quad \quad \quad i,j = 1, \ldots, n+1 ,
\end{equation}
where $\tau_1, \ldots, \tau_n$ are $n$ additional commuting variables.
The definition of $H$ is the same as the definition of $G$, except that
different variables are used.  Reasoning similar to that in the original
case gives the following expression for $z_n$.
\begin{equation}
z_n = T \, (\log FGH)_{1, n+1} \, ,
\label{b}
\end{equation}
where the definition of the $T$ operator now has been extended to also
putting a $w$ at the $i$-th position of a product of $x$'s, $y$'s and $w$'s
for an occurrence of $\tau_i$.  For example, when $n$ is 4, we have
$T\, (\sigma_2 \, \tau_3) = xywx$.  The results obtained from Eq.~(\ref{b})
agree with those obtained from Reutenauer's generalization\cite{Reutenauer}
of Goldberg's theorem.

An example of how these methods can be used with Mathematica is shown
below.  For ease of coding, the notation has been changed slightly.

\begin{verbatim}
p3[n_] := p3[n] = (
           F = Table[1/(j-i)! Product[s[k,"x"],{k,i,j-1}],{i,n+1},{j,n+1}];
           G = Table[1/(j-i)! Product[s[k,"y"],{k,i,j-1}],{i,n+1},{j,n+1}];
           H = Table[1/(j-i)! Product[s[k,"w"],{k,i,j-1}],{i,n+1},{j,n+1}];
           qthpower = IdentityMatrix[n+1]; FGm1 = F.G.H - qthpower; Expand[
           -Sum[qthpower=qthpower.FGm1; (-1)^q / q qthpower,{q,n}][[1,n+1]]])

translated3[n_] := (temp = p3[n]; Sum[term = Apply[List, temp[[i]]]; term[[1]]*
           Apply[StringJoin, Take[term,-n] /. s[j_,k_]->k], {i,Length[temp]}])

In[3]:= translated3[2]

        -wx   wy   xw   xy   yw   yx
Out[3]= --- - -- + -- + -- + -- - --
         2    2    2    2    2    2
\end{verbatim}

\section{conclusion}

The results contained in this paper provide a means of computing the
entire $n$-th order term in the Baker-Campbell-Hausdorff series,
without the use of noncommuting variables, sums over partitions,
or other complicated operations.  One application is in writing
simple programs in standard computer languages to calculate the
Baker-Campbell-Hausdorff series.  Such programs do not result in
a significant reduction in computer time needed.  Rather, the
programming involved is simplified.  The sample program included
in this paper can be shortened to just a few lines, and it is not
necessary to load special software packages.  The calculation of 
higher-order terms in the Baker-Campbell-Hausdorff series is usually
done in computer languages that do not have symbol manipulation,
because they are faster.  This paper also explains how a simple
program can be written in such a language to calculate the
series.

This paper does not address the question of expressing the
Baker-Campbell-Hausdorff series in terms of commutators.  As
explained in Sec.~\ref{thm}, if an expression in terms of
commutators is desired, the substitution due to
Dynkin\cite{Oteo,Dynkin,T,Bose} may be used to transform
the results calculated here.

Future research could include finding alternative ways to calculate
certain sequences of elements in graded free Lie algebras, such as
those in Ref.~\onlinecite{MKO} (which also contains an interesting
method of computing the Baker-Campbell-Hausdorff series by numerically
integrating a differential equation).  These quantities occur in the
optimization of numerical algorithms involving computations in Lie
algebras.  Graded Lie algebra bases can be used in the construction of
Runge-Kutta methods on manifolds.

\end{document}